\documentstyle[12pt]{article}

\def\p{{\rm {\bf p}}}
\def\q{{\rm {\bf q}}}

\def\exp{{\rm exp}}
\def\g{\mbox{\boldmath$\gamma$}}

\begin{document}

\vspace{1.5cm}

\begin{center}
{\bf COLLECTIVE EXCITATIONS OF MASSIVE DIRAC PARTICLES IN HOT AND DENSE
MEDIUM}

\vspace{1.5cm}

{\bf O.K.Kalashnikov}
\footnote{Permanent address: Department of
Theoretical Physics, P.N.Lebedev Physical Institute, Russian
Academy of Sciences, 117924 Moscow, Russia. E-mail address:
kalash@td.lpi.ac.ru}

High-Energy Physics

ICTP, Trieste

\vspace{2.5cm}

{\bf Abstract}
\end{center}

The one-loop dispersion equation which defines the collective
excitations of the massive Dirac particles in hot and dense quark-gluon
medium is obtained in the high temperature limit for the case $m<<T$
and solved explicitly for all $|\q|$ when $\mu=0$.  Four
well-separated spectrum branches (quasi-particle and quasi-hole
excitations) are found and their behaviors for the small and large
$|\q|$ are investigated.  All calculations are performed using the
temperature Green function technique and fixing the Feynman gauge.
The gauge dependency of the spectra found are briefly discussed.

\newpage

\section{Introduction}
The studying of the collective excitations in hot and dense medium is a
very actual problem for the current physics and for the chromodynamics
in the first rate. In the medium all particles (fermions the same as
bosons) lose their individual properties and the collective excitations
arise  which (unlike the ordinary vacuum physics at $T,\mu=0$) have
many new peculiarities. Namely these collective excitations determine
the bulk of kinetical and thermodynamical properties of hot and dense
medium and are very important for many processes taken place, for
example, inside the quark-gluon medium.  Moreover, inside the
quark-gluon medium (when $\mu$ or $T$ are nonzero) the new collective
(hole) excitations of fermions arise [1,2] which are different from the
quasi-particle ones and their peculiarities (e.g. the minimum of the
quasi-hole branches at the finite momentum and the "wrong" relation
between chirality and helicity) can be exploited for searching the new
physical consequences.  All these collective modes due to the medium
have the effective masses (the same as the plasmon masses for
gluons [3]) which independently from the bare masses are generated
dynamically and these dynamical masses are not small for the large
$T,\mu$-parameters. In particular, for the initially massive Dirac
particles there is the set of four effective masses [4,5,6] which,
in a general case, are well-separated and are always nonzero in the
medium.

The goal of this paper is to present the one-loop dispersion equation
which defines the collective excitations of the massive Dirac particles
in hot and dense quark-gluon plasma in the high temperature limit for
the case $m<<T$ and to solve it explicitly for all $|\q|$ when $\mu=0$.
We use the standard temperature Green function technique and fix the
Feynman gauge for explicit calculations.  The case of a zero damping is
only considered and many additional problems connected with calculating
the damping rate [7] are not discussed. Four well-separated spectrum
branches are established and their behaviors for the small and large
$|\q|$ are investigated. The gauge dependency of the spectra found are
briefly discussed. To start we choose hot and dense QCD although many
results are model independently.

\section{ QCD Lagrangian and  quark self-energy}

The QCD Lagrangian in covariant gauges has the form
\setcounter{equation}{0}
\begin{eqnarray}
{\cal L}=&-&\frac{1}{4}{G_{\mu\nu}^a}^2+N_f{\bar \psi}
[\gamma_{\mu}(\partial_{\mu}-\frac{1}{2}
ig\lambda^aV_{\mu}^a)+m]\psi \nonumber\\ &-&\mu N_f{\bar
\psi}\gamma_4\psi +\frac{1}{2\alpha}(\partial_{\mu}V_{\mu}^a)^2 +{\bar
C}^a (\partial_{\mu}\delta^{ab}+gf^{abc}V_{\mu}^c)\partial_{\mu}C^b
\end{eqnarray}
where $G_{\mu\nu}^a=\partial_{\mu}V_{\nu}^a-\partial_{\nu}V_{\mu}^a
+gf^{abc}V_{\mu}^bV_{\nu}^c$  is the Yang-Mills field strength;
$V_{\mu}$ is a non-Abelian gauge field; $\psi$(and ${\bar
 \psi}$) are the quark fields in the SU(N)-fundamental representation
($\frac{1}{2}\lambda^a $ are its generators and $f^{abc}$ are the
SU(N)-structure constants) and $C^a$ (and ${\bar C}^a$) are the ghost
Fermi fields. In Eq.(1) $\mu$ and $m$ are the quark chemical potential
and the bare quark mass, respectively, $N_f$ is the number of quark
flavours and $\alpha$ is the gauge fixing parameter ($\alpha=1$ for
the Feynman gauge). The metric is chosen to be Euclidean and
$\gamma_{\mu}^2=1$.
We use the exact Schwinger-Dyson equation for the temperature quark
Green function
\begin{eqnarray} G^{-1}(q)=G_0^{-1}(q)+\Sigma(q)
\end{eqnarray}
where the quark self-energy in any gauge has the simple representation
[8]
\begin{eqnarray}
\Sigma(q)=\frac{N^2-1}{2N}\frac{g^2}{\beta}\sum_{p_4}^F
\int\frac{d^3p}{(2\pi)^3}{\cal D}_{\mu\nu}(p-q)\gamma_{\mu}G(p)
\Gamma_{\nu}(p,q|p-q)\;.
\end{eqnarray}
The representation (3) is exact but we calculate $\Sigma(q)$ only in
the one-loop approximation using the bare Green functions in Eq.(3) and
fixing the Feynman gauge (i.e.  taken the appropriate ${\cal
D}$-function). All ultraviolet divergencies are renormalized as usual
but the infrared ones (which also arise in the high temperature
expansion when $m<<T$) will be eliminated phenomenologically.

At first the summation over the spinor indices is performed in Eq.(3)
using the standard $\gamma$-matrix algebra
\begin{eqnarray}
\Sigma(q)
=\frac{N^2-1}{N}\frac{g^2}{\beta}\sum_{p_4}^F\int\frac{d^3p}
{(2\pi)^3}\;\frac{i\gamma_{\mu}{\hat p_{\mu}}+2m}{({\hat
p^2}+m^2)\;(p-q)^2}
\end{eqnarray}
and then the summation is performed over the Fermi frequencies
$p_4=2\pi T(n+1/2)$ using the well-known prescription [8]. Here
${\hat p}=\{(p_4+i\mu),\p \}$ is the convenient abbreviation for
vectors with $\mu$. All terms found are collected in the convenient
form using the simple algebraic trasformations and the final result is
given by

\begin{eqnarray}
&&\!\!\!\!\!\!\!\!\!\!\!\!\!\!\!\Sigma(q)=
-\frac{g^2(N^2-1)}{N}\int\frac{d^3p}{2(2\pi)^3}\;\left\{\;\Bigr[
\frac{1}{\epsilon_\p}\;\frac{n_\p^+\;[\gamma_4\epsilon_\p+(i\g\p+2m)]}
{[q_4+i(\mu+\epsilon_\p)\;]^2+(\q-\p)^2}\right.\nonumber\\
&&\!\!\!\!\!\!\!\!\!\!\!\!\!\!\!\!\!\! \left.+\;\frac{n_\p^B}{|\p|}
\;\frac{(|\p|+\mu-iq_4) \gamma_4-[i\g(\q-\p)+2m]}{[q_4+
i(\mu+|\p|)\;]^2+\epsilon_{\p-\q}^2}\;\Bigr]
-\Big[h.c.(m,\mu)\rightarrow-(m,\mu)\Big]\right\}
\end{eqnarray}
where $\epsilon_\p=\sqrt{\p^2+m^2}$ is the bare quark energy;
$n_\p^{B}=\left\{\exp\beta|\p|-1\right\}^{-1}$ and $n_\p^{\pm}=
\left\{\exp\beta(\;\epsilon_\p \pm \mu)+1\right\}^{-1}$ are the Bose
and Fermi occupation numbers, respectively.

For futher calculations it is convenient to introduce two new
functions and to rewrite Eq.(5) as follows
\begin {eqnarray}
\Sigma(q)=i\gamma_{\mu}K_{\mu}(q)+m\;Z(q)
\end{eqnarray}
where $K_\mu(q)=q_\mu a(q)+iu_\mu b(q)$ and $u_\mu=\{1,0\}$ is
the unit medium vector. All functions separately depend on $q_4$ an
$|\q|$ as usual in the medium case. Eq.(6) presents the one-loop
decomposition of $\Sigma(q)$ which, however, is not complete here
(see [9] for detail) since a number of functions are generated only in
the multi-loop calculations.  Using the decomposition (6) we transform
Eq.(2) into the form
\begin{eqnarray}
G(q)=\frac{-i\gamma_{\mu}({\hat q_{\mu}}+K_{\mu})+m\;(1+Z)}
{({\hat q_{\mu}}+K_\mu)^2\;+\;m^2\;(1+Z)^2}
\end{eqnarray}
which gives the correct nonperturbative structure for this function.
Setting up the determinant of Eq.(7) to be zero, we find the dispersion
equation
\begin{eqnarray}
({\hat q_{\mu}}+K_{\mu})^2\;+\,m^2\;(1+Z)^2=0
\end{eqnarray}
which defines the collective excitation spectra after
the standard analytic continuation.

\section{ Collective excitations in the high temperature limit}
Here we use Eq.(8) to find the dispersion equation for the
collective excitations of the massive Dirac particles in hot and dense
quark-gluon plasma when $m<<T$. The different limits of this equation
are discussed and it is solved exactly for the massive fermion case
with $\mu=0$.  The spectrum branches are found for all $|\q|$ and their
limits for the small and large momenta are presented explicitly. The
case of a zero damping is only considered and due to this fact our
analytical continuation is trivial.

Our starting point is the dispersion equation (8)
\begin {eqnarray}
[\;(iq_4-\mu)-{\bar K}_4]^2\;=\;\q^2\;(1+K)^2 +m^2(1+Z)^2
\end{eqnarray}
with $m\ne 0$ and we use Eq.(5) to find its high temperature expansion
when $m<<T$. Here $K_4=i{\bar K}_4$ and we take into account only the
leading $T^2$-terms with the $\mu/T$-corrections within Eq.(9).  In
this case all functions which define Eq.(9) can be simplified as
follows
\begin {eqnarray}
&&K(q_4,\q)\;=\;\frac{I_K}{\q^2}\Bigr(\;1+\frac{\xi}{2}\ln
\frac{\xi-1}{\xi+1}\Bigr)\;+\;I_B\;\Bigr(\;\xi-\frac{1}{2}
(1-\xi^2)\ln\frac{\xi-1}{\xi+1}\;\Bigr)
\end{eqnarray}
\begin {eqnarray}
-{\bar K}_4(q_4,\q)\;=\;\frac{I_K}{2|\q|}
\ln\frac{\xi-1}{\xi+1} +I_B \,,\qquad
-Z(q_4,\q)\;=\;2I_Z\;+\;\frac{2I_B}{|\q|}
\ln\frac{\xi-1}{\xi+1}
\end{eqnarray}
that gives a possibility to solve Eq.(9) explicitly. Here
$\xi=\omega/|\q|$ is a convenient variable and the integrals are
defined to be
\begin {eqnarray}
&&I_K\;=\;\frac{g^2(N^2-1)}{N}\int\limits_0^{\infty}\frac{d|\p|}{4\pi^2}
\;|\p|\;\Bigr[\;\frac{n_\p^++n_\p^-}{2}\;+\;n_\p^B\;\Bigr] \\
&&I_B=-\frac{g^2(N^2-1)}{N}\int\limits_0^{\infty}
\frac{d|\p|}{8\pi^2}\;\frac{n_\p^+-n_\p^-}{2} \\
&&I_Z=\frac{g^2(N^2-1)}{N}\int\limits_0^{\infty}
\frac{d|\p|}{8\pi^2}\;\frac{n_\p^++n_\p^-}{2\epsilon_\p}\;.
\end{eqnarray}
The integral $I_Z$, however,  is redefined to avoid the infrared
divergencies which arise after the high temperature expansion has been
performed for $Z(q_4,|\q|)$
\begin{eqnarray}
&&Z(q)=-\frac{g^2(N^2-1)}{N}\int\frac{d^3p}
{(2\pi)^3}\;\left\{\;\Bigr[\;\frac{1}{\epsilon_\p}\;\frac{n_\p^+}
{[q_4+i(\mu+\epsilon_\p)\;]^2+(\q-\p)^2}\right.\nonumber\\
&&-\left.\frac{n_\p^B}{|\p|}\;\;\frac{1}
{[q_4+i(\mu+|\p|)\;]^2+\epsilon_{\p-\q}^2}\;
\Bigr]\;+\;\Big[\;h.c.(\mu\rightarrow-\mu)\;\Bigr]\;\right\}\;.
\end{eqnarray}
The last expression is extracted from Eq.(5).

Now one should plug the expressions found above into Eq.(9) and perform
a number of the algebraic transformations to find $\omega=\xi|\q|$.
Here $\omega=(iq_4-\mu)$ and the variable $\xi$ is more convenient than
$|\q|$. The result is the equation of the fourth power with respect to
$\omega(\xi)$
\begin {eqnarray}
&&\omega^4\Bigr[\xi^2-(1+b(\xi)I_B)^2\;\Bigr]+2\omega^3\;\xi^2\;I_B
+\omega^2\;\xi^2\Bigr[\;I_B^2-m_R^2+2\;d(\xi)I_K\nonumber\\
&&-2(1+b(\xi)I_B)(1+d(\xi))I_K\;\Bigr]+
2\omega\;\xi^2\;d(\xi)\;I_B\;[\;I_K+4m_R^2\;]\nonumber\\
&&+I_K^2\;\xi^2\;[\; d(\xi)^2-\xi^2\;(1+d(\xi))^2\;]-
16m^2\;\xi^2\;d(\xi)^2I_B^2\;=0
\end{eqnarray}
where $m_R=m(1-2I_Z)$ is the renormalized fermionic mass and functions
$d(\xi)$ and $b(\xi)$ are given by
\begin {eqnarray}
&&d(\xi)=\frac{\xi}{2}\;\ln\frac{\xi-1}{\xi+1}\nonumber\\
&&b(\xi)=\xi-\frac{1}{2}(1-\xi^2)\ln\frac{\xi-1}{\xi+1}\;.
\end{eqnarray}
The obtained dispersion equation being very complicated is not solved
exactly. However in the long wavelength limit (when
$\xi\rightarrow\infty$) it can be simplified as follows
\begin {eqnarray}
\Bigr[\omega^2+\omega(I_B-\eta m_R)-(I_K+4\eta mI_B)\Bigr]\cdot
\Bigr[\omega^2+\omega(I_B+\eta m_R)-(I_K-4\eta mI_B)\Bigr]=0
\end{eqnarray}
and one finds a rather simple solution [6]
\begin {eqnarray}
\omega(0)=\frac{1}{2}\Big[\eta\;m_R-I_B\Big]\pm
\sqrt{\;\frac{[\eta\;m_R-I_B]^2}{4} +(I_K+4\eta mI_B)}
\end{eqnarray}
which demonstrates four well-separated effective masses:
two of them are related to the quasi-particle excitations and other
two present the quasi-hole ones.  Here $\eta=\pm 1$ and the parameters
$m$ and $\mu$ are nonzero.

The solutions for all $|\q|$ are possible to find within Eq.(16) if
either $m$ or $\mu$-parameters are equal to zero.

The case $m=0$ with $\mu\ne 0$ has been recently considered in
detail and the result has the form [6]
\begin{eqnarray}
E(\xi)=\mu-\frac{\xi\;I_B}{2(\xi-\eta)}\;
\pm\sqrt{\frac{\xi^2I_B^2}{4(\xi-\eta)^2}+I_K\xi^2\Bigr(\;\frac{\eta}
{\xi-\eta}+\frac{\eta}{2}\ln\frac{\xi-1}{\xi+1}\;\Bigr)}
\end{eqnarray}
which extends the well-known result found in [1,2] to the case $\mu\ne
0$. Here we restore the physical variable $E=ip_4$. The variable $\xi$
runs in $1<\xi<\infty$ and the long wavelength limit corresponds to
$\xi\rightarrow\infty$. For this limit one finds the very simple result
\begin{eqnarray}
E(0)=\mu-\frac{I_B}{2}\;\pm\;\sqrt{\frac{I_B^2}{4}+I_K}
\end{eqnarray}
which can be compared with the interpolation formula in [10].

Another possibility is the case $m\ne 0$ but with
$\mu=0$ when Eq.(16) can be solved exactly for all $|\q|$ as well.
Namely this possibility is the subject of this paper and will be
discussed below when $m<<T$.  Now $I_B=0$ and keeping the proper
accuracy of calculations the solution of Eq.(16) is found to be
\begin{eqnarray}
\omega_{\pm}(\xi)^2=\frac{\xi^2(\;2I_K+m_R^2)}{2(\xi^2-1)}\;
\pm\;\sqrt{\frac{\xi^4}{(\xi^2-1)^2}\Bigr[\;(b(\xi)I_K)^2
+\;m_R^2(\;I_K+m_R^2/4)\Bigr]}\;.
\end{eqnarray}
These spectra are our main result. They present the collective
excitations of massive Dirac particles in hot medium for all $|\q|$
when $m<<T$. Two spectrum branches (when the sign is plus) correspond
to the quasi-particle excitations and other two (when the sign is
minus) are the quasi-hole ones. These spectrum branches have a rather
different asymptotical behaviors and many other different properties.

The long wavelength behaviour of these spectra (when
$\xi\rightarrow\infty$) has the form
\begin {eqnarray}
\omega_{\pm}(|\q|)^2=\;M_{\pm}^2\;+\;\Bigr(\;M_{\pm}^2\pm
\frac{4}{9}\frac{I_K^2}{\sqrt{\;m_R^2(m_R^2+4I_K\;)}}
\;\Bigr)\;\frac{|\q|^2}{M_{\pm}^2}\;+\;O(|\q|^4)
\end{eqnarray}
where the effective masses squared are given by
\begin {eqnarray}
M_{\pm}^2=\;\frac{m_R^2}{2}+I_K\pm
\sqrt{\;m_R^2\Bigr(\frac{m_R^2}{4}+I_K\;\Bigr)}\;.
\end{eqnarray}
These masses are different for four spectrum branches
$M_{\pm}=\frac{1}{2}(\eta m_R\pm \sqrt{m_R^2+4I_K})$ and are in
agreement with the results [4,5]. Here $\eta=\pm 1$.

However, this is not the case when the second term in Eq.(23) is taken
into account.  This term is not in agreement with one obtained in
[4,5]. Although it qualitatively coincides with the result presented in
[5] but there is the essential difference with [4] where the linear
term was found. It is also important that the quasi-hole spectra
$\omega_-(|\q|)^2$ are very sensitive to the choice of
$m,T$-parameters.  In many cases these spectra are the monotonical
functions for the small $|\q|^2$ and the well-known minimum [1]
disappears. This minimum always exists only for the massless particles
but when $m\ne 0$ the special conditions are necessary to generate it.

In the high momentum region the asymptotical behaviors found for the
quasi-particles and the quasi-hole excitations are unlike completely.
The quasi-particle spectrum branches are approximated as follows
\begin{eqnarray}
\omega_{+}(|\q|)^2=\;|\q|^2\;+\;(2I_K+m_R^2)\;
-\;\frac{I_K^2}{|\q|^2}\ln\frac{4|\q|^2}{2I_K+m_R^2}
\end{eqnarray}
where the nonanalytical term is unessential. Another situation takes
place for the quasi-hole excitations which do not exist in the vacuum
(when $T$ and $\mu$ are equal to zero). They very fast disappear and
their asymptotical behaviour is found to be
\begin{eqnarray}
\omega_{-}(|\q|)^2=
\;|\q|^2\;+4|\q|^2\exp(-\frac{|\q|^2(2I_K+m_R^2)}{I_K^2})\;.
\end{eqnarray}
In the high momentum region these spectrum branches approach to the line
$\omega^2=|\q|^2$ more quickly than (25).

\section{Conclusion}
To summarize we have obtained and solved the one-loop dispersion
equation for the massive fermions at finite temperature. Our solution
gives the collective Fermi excitations for all $|\q|$ and we establish
that they have four well-separated branches: two of them present the
quasi-particle excitations and two other correspond to the quasi-hole
ones. The splitting calculated demonstrates that the effective masses
for all branches are different when $m\ne 0$ and these masses are
always nonzero in the medium. The asymptotical behavior found for the
small $|\q|$ shows that the difference between the initially massive
and massless fermions is kept although the dynamical mass is always
generated and all their collective excitations are massive.  For the
massless fermions one finds that the spectrum minimum always exist and
the leading asymptotical term for the small $|\q|$ is linear.  However
this is not the case for the initially massive fermions.  When $m\ne 0$
the spectrum minimum, as rule, disappears as well as the linear term
and the term $|\q|^2$ gives the leading asymptotic behavior for the
small $|\q|$. The gauge invariance of the results found, unfortunately,
is not proved and there are not any guaranties that this is, indeed,
so. Here the situation is completely unclear and there is only the fact
that the dynamical mass for the case $m,\mu=0$ is the gauge invariant
object.  All other quantities are gauge dependent, in any case, within
the one-loop calculations. Of course it is not excluded that the
Braaten-Pisarski resummation is necessary to improve this situation but
this question is not so evident as for the usual damping rate
calculations.

\begin{center}
{\bf Acknowledgments}
\end {center}
I would like to thank to S. Randjbar-Daemi for the invitation me to the
International Centre for Theoretical Physics in Trieste and to all the
colleagues of this center for the kind hospitality.

\begin{center}
{\bf References}
\end{center}

\renewcommand{\labelenumi}{\arabic{enumi}.)}
\begin{enumerate}

\item{ V.~V.~Klimov, Yad.Fiz. {\bf 33} (1981) 1734 (Sov. J. Nucl. Phys.
{\bf 33} (1981) 934); Zh. Eksp. Teor. Fiz. {\bf 82} (1982) 336 (Sov.
Phys. JETP  {\bf 55} (1982) 199).}

\item{ H.~A.~Weldon, Phys. Rev. {\bf D26} (1982) 2789.}

\item{ O.~K.~Kalashnikov and V.~V.~Klimov, Yad. Fiz. {\bf 31} (1980)
1357 (Soviet J. Nucl. Phys. {\bf 31} (1980) 699).}

\item{ R.~D.~Pisarski, Nucl. Phys. {\bf A498} (1989) 423c.}

\item{ C.~Quimbay and ~S.~Vargas-Castrillon, Nucl. Phys. {\bf B451}
(1995) 265.}

\item{ O.~K.~Kalashnikov, Mod. Phys. Lett. {\bf A12} (1997) 347.}

\item{ Jean-Paul Blaizot and E.Iancu, Phys. Rev. {\bf D55} (1997) 973.}

\item{ E.~S.~Fradkin, Proc. (Trudy ) P.N.Lebedev Physics Inst. {\bf 29}
(1967) 1.}

\item{ O.~K.~Kalashnikov, Pis'ma  Zh. Eksp. Teor. Fiz. {\bf 41} (1985)
477;\protect\\ (JETP Lett.  {\bf 41} (1985) 582).}

\item{ K.~Kajantie and P.~V.~Ruuskaven, Phys.Lett. {\bf B 121 }
(1983) 352.}

\end{enumerate}

\end{document}